\documentclass[preprint,12pt,3p]{elsarticle}
\usepackage[bookmarks=false]{hyperref}
\usepackage{amssymb}
\usepackage{amsmath}
\usepackage{bm}
\usepackage{color} 
\usepackage[normalem]{ulem} 
\usepackage{mathtools}
\usepackage{placeins}
\usepackage{caption}
\usepackage{graphicx}
\usepackage{lineno}

\makeatletter
\def\ps@pprintTitle{%
   \let\@oddhead\@empty
   \let\@evenhead\@empty
   \let\@oddfoot\@empty
   \let\@evenfoot\@oddfoot
}
\makeatother
\begin{document}
\begin{frontmatter}


\title{Neutron Spectroscopy for Pulsed Beams with Frame Overlap using a Double Time-of-Flight Technique}


\ead{bethany@nuc.berkeley.edu}
\cortext[cor1]{Corresponding author.}

\author[1]{K.~P.~Harrig}
\author[1]{B.~L.~Goldblum\corref{cor1}}
\author[1]{J.~A.~Brown}
\author[3]{D.~L.~Bleuel}
\author[1,2]{L.~A.~Bernstein}
\author[1]{J.~Bevins}
\author[1]{M.~Harasty}
\author[1]{T.~A.~Laplace}
\author[1]{E.~F.~Matthews}


\address[1]{Department of Nuclear Engineering, University of California, Berkeley, California 94720, USA}
\address[3]{Lawrence Livermore National Laboratory, Livermore, California 94550, USA}
\address[2]{Lawrence Berkeley National Laboratory, Berkeley, California, 94720, USA}


\begin{abstract}
A new double time-of-flight (dTOF) neutron spectroscopy technique has been developed for pulsed broad spectrum sources with a duty cycle that results in frame overlap, where fast neutrons from a given pulse overtake slower neutrons from previous pulses. Using a tunable beam at the 88-Inch Cyclotron at Lawrence Berkeley National Laboratory, neutrons were produced via thick-target breakup of 16~MeV deuterons on a beryllium target in the cyclotron vault. The breakup spectral shape was deduced from a dTOF measurement using an array of EJ-309 organic liquid scintillators. Simulation of the neutron detection efficiency of the scintillator array was performed using both GEANT4 and MCNP6. The efficiency-corrected spectral shape was normalized using a foil activation technique to obtain the energy-dependent flux of the neutron beam at zero degrees with respect to the incoming deuteron beam. The dTOF neutron spectrum was compared to spectra obtained using HEPROW and GRAVEL pulse height spectrum unfolding techniques. While the unfolding and dTOF results exhibit some discrepancies in shape, the integrated flux values agree within two standard deviations. This method obviates neutron time-of-flight spectroscopy challenges posed by pulsed beams with frame overlap and opens new opportunities for pulsed white neutron source facilities.

\end{abstract}



\end{frontmatter}


\section{Introduction}
\label{intro}

Fast neutron spectroscopy has been performed using a wide range of techniques, including time-of-flight (TOF) approaches \cite{brooks,meulders,bleuel}, proton recoil spectrometry \cite{donzella,bame,geiger}, and pulse height and activation foil unfolding methods \cite{lawrence,maeda,sen,reginatto}. This work focuses on enabling neutron TOF spectroscopy, where the time of travel of the neutron is measured over a fixed distance, in previously inaccessible regimes. For broad spectrum pulsed neutron sources, frame overlap can result in ambiguous TOF signals where fast neutrons from a given pulse have the same observable TOF as slower neutrons from previous pulses. As a result, neutron TOF experiments with pulsed sources have traditionally been limited by the duty cycle of the source \cite{schober,cierjacks}.

Approaches to enabling TOF spectroscopy in pulsed beam scenarios have largely focused on physically filtering the neutron energy range, including the use of mechanical beam choppers \cite{fermi} and electrostatic deflectors \cite{abrahamsson}. In addition to these filters, many pulsed source TOF measurements are forced to implement low-energy software thresholds to eliminate ``wrap-around" effects, which can greatly limit low-energy spectral data \cite{weaver}. Some approaches use shorter flight paths to detect the full energy range, thereby increasing the overall uncertainty of the TOF measurement and decreasing the energy resolution \cite{saltmarsh}. 

Building on the work of Schweimer~\cite{schweimer}, a new method for neutron TOF spectroscopy tailored for pulsed sources with frame overlap has been developed. Neutron-proton elastic scattering kinematics is exploited using a double TOF (dTOF) technique that allows for pulse association of temporally ambiguous neutron TOF signals. This technique is demonstrated using a pulsed deuteron breakup neutron source at the 88-Inch Cyclotron at Lawrence Berkeley National Laboratory (LBNL). Descriptions of the facility, experimental setup, electronics, and materials are provided in Sec.~\ref{exp}. Section~\ref{beam_char} details the dTOF approach, efficiency simulations of the scintillator array, and foil activation analysis methods. The measured neutron energy spectrum is provided in Sec.~\ref{results-yo} and compared to results obtained using pulse height spectrum unfolding methods. Concluding remarks are presented in Sec.~\ref{conc}.
\section{Experimental setup}
\label{exp}

\begin{figure}
\centering
\includegraphics[width=0.7\textwidth]{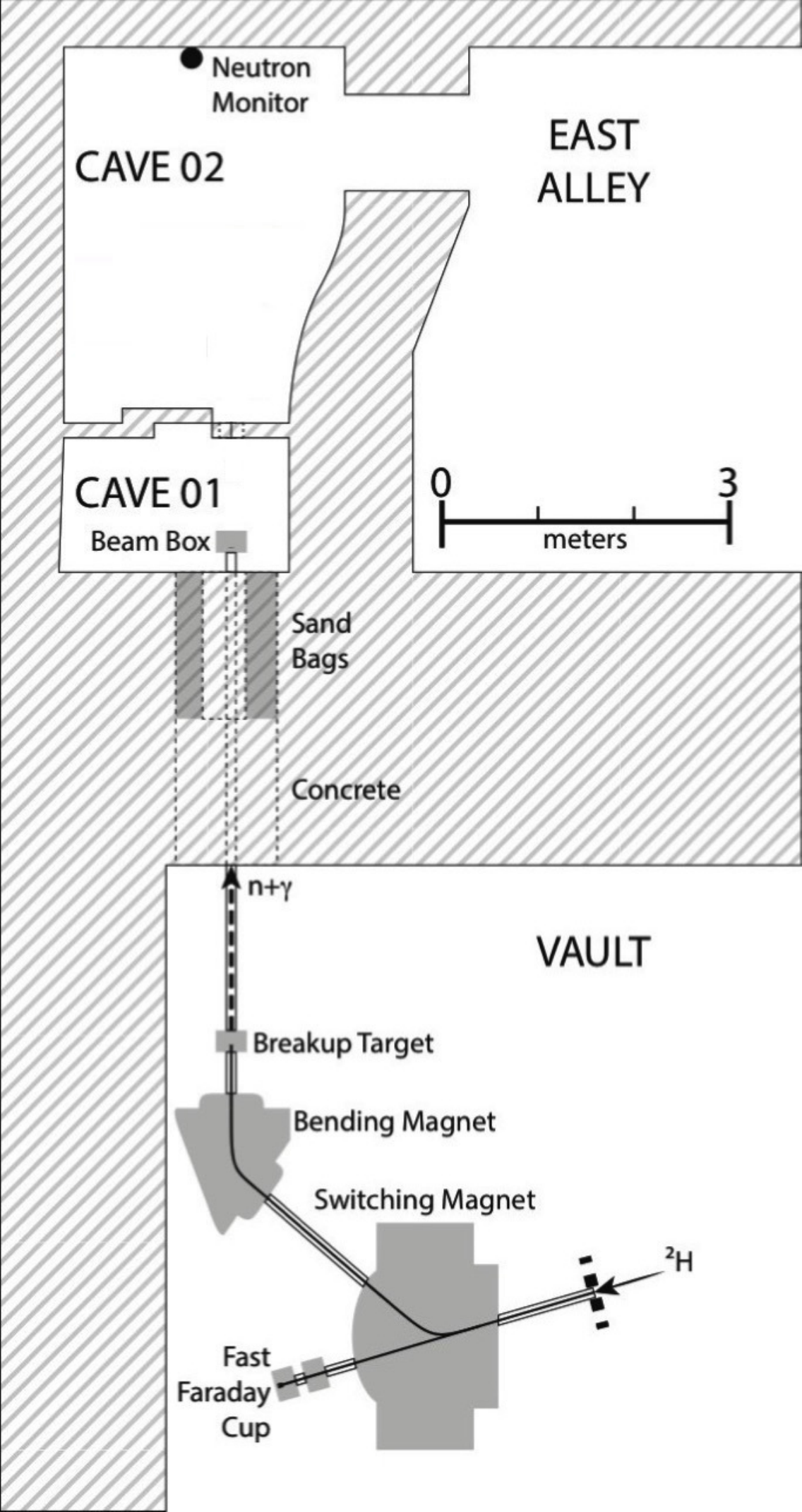}
\caption{Schematic representation of the 88-Inch Cyclotron vault and beam line to Cave~0. The Cave~0 experimental endstation is comprised of two enclosures, Cave~01 and Cave~02, separated by a lead-lined door outfitted with a beam port.}
\label{explayout}
\end{figure}

The 88-Inch Cyclotron at LBNL is a variable energy, high-current, multi-particle cyclotron capable of accelerating deuterons up to a maximum energy of 65~MeV with maximum currents on the order of 10 particle-$\mu$A. In this work, a $^2$H$^+$ beam was accelerated to 16~MeV at a current of $\sim$180~nA during the neutron scattering measurements, $\sim$0.66 nA during the neutron singles pulse height measurements, and $\sim$1.8~$\mu$A during the foil irradiation. The deuteron beam was directed along the Cave~0 beam line and optically aligned using a phosphor located in the Cave~01 beam box, as shown in Fig.~\ref{explayout}. A Faraday cup located inside the cyclotron vault was equipped with a 3-mm-thick beryllium breakup target with a tantalum backing and plunged along the Cave~0 beam line \cite{bleuel,mcmahan}. The beryllium target is embedded in a 3.8-mm-thick electrically isolated tantalum disk and backed by a 14.5-mm-thick copper cooling assembly. The resulting neutrons and photons entering the experimental area were collimated by $\sim$3~m of concrete and $\sim$1.5~m of sand bags encasing the beam pipe, producing an open-air neutron beam in the experimental area.

As the 88-Inch Cyclotron is a pulsed ion source, the 16~MeV $^2$H$^+$ beam corresponds to a cyclotron radio frequency of $6.31$~MHz and a pulse period of $158.5$~ns. The standard deviation of the temporal profile of the incoming beam pulse was approximately 6~ns. For the 16~MeV beam with a flight path of 6.84~m, for instance, neutrons with energies less than $9.87$~MeV (i.e., TOF $> 158.5$~ns) overlap with neutrons and gamma rays from the subsequent beam pulse. 

\begin{figure}
\centering
\includegraphics[width=0.7\textwidth]{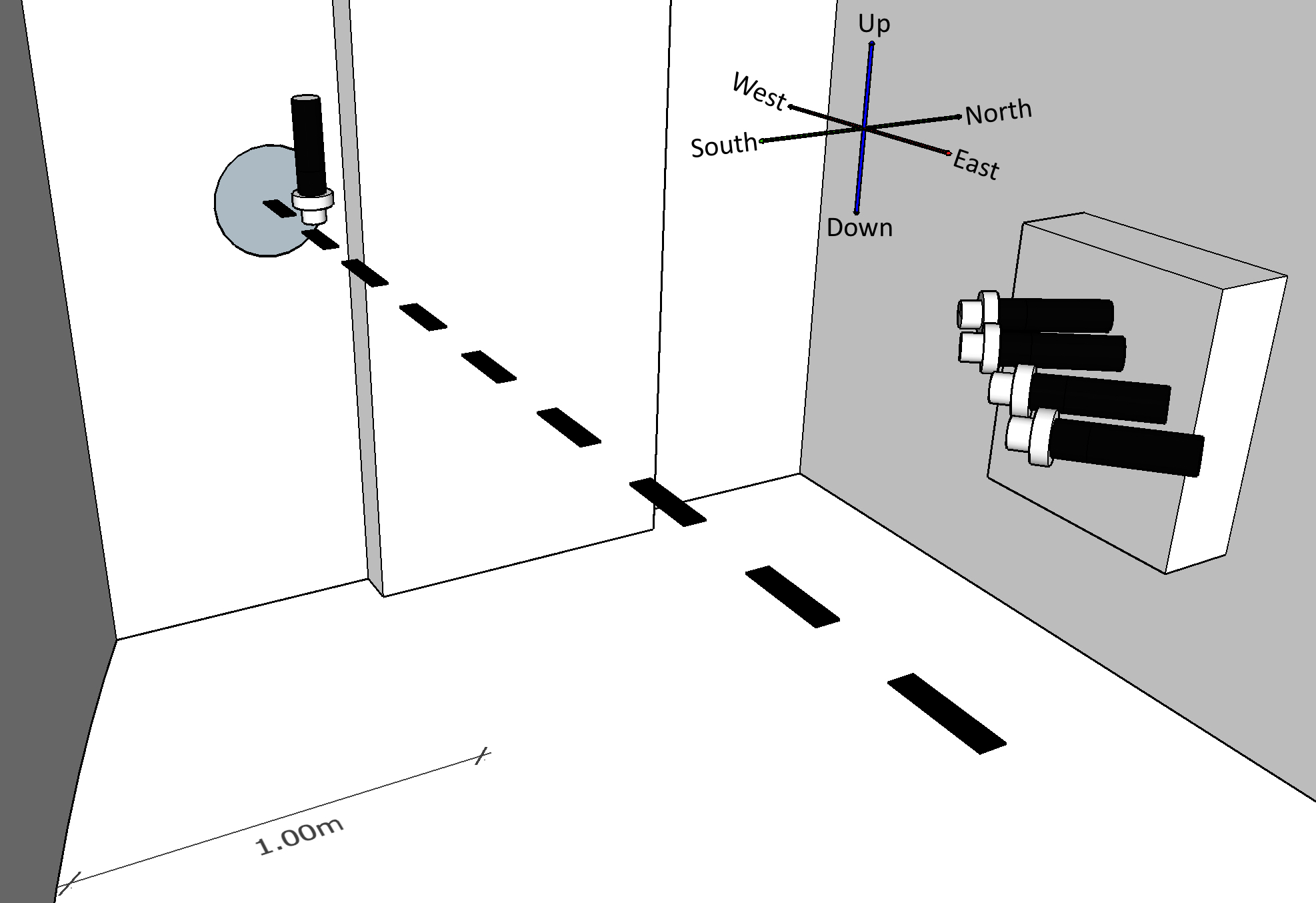}
\caption{Schematic view of the scintillator array in the Cave~02 experimental area. The black dashed line indicates beam line center. \label{layout}}
\end{figure}

The dTOF measurements were performed using an array of pulse-shape-discriminating organic liquid scintillators. The scintillator array was composed of one cell in beam (i.e., target cell) and four cells out of beam (i.e., scatter cells), illustrated in Fig.~\ref{layout}. The cells consisted of cylindrical Hamamatsu H1949-50 photomultiplier tubes (PMTs) biased from ${-1.75}$ to ${-2.1}$ kV, coupled to cylindrical liquid organic EJ-309 scintillators, and were positioned using adjustable camera tripods. All scintillators had a height of 5.08~cm, a diameter of 5.08~cm, and were enclosed by an aluminum casing.

\begin{table*}
\centering
\begin{tabular}{cccccc}
\hline
Detector & $x$ (cm) & $y$ (cm) & $z$ (cm) & $L$ (cm) & $\phi$ (degrees) \\
\hline
Target    & 36.3  & 0     &3.5& 683.8& -     \\
Scatter 0 & 154.1 & 99.5 & 0 & 154.2 & 40.2 \\
Scatter 1 & 169.3 & 86.2 & 0 & 158.5 & 33.0 \\
Scatter 2 & 187.7 & 76.6 & 0 & 169.7 & 26.9 \\
Scatter 3 & 204.5 & 65.0 & 0 & 180.3 & 21.1 \\
\hline
 \end{tabular}
 \caption{Target and scatter cell configuration. The $x$-coordinate was determined relative to the west wall, the $y$-coordinate relative to beam line center, and the $z$-coordinate relative to the plane of the beam. Each coordinate was measured using a cross-line laser level. The flight path, $L$, is defined as the distance from the breakup target to the center of the target cell and the distance from the center of the target cell to the center of the scatter cells for the target and scatter cells, respectively. The angle, $\phi$, was determined relative to the axis defined by the incoming beam. \label{config}}
 \end{table*}

The flat aluminum surface of the target cell, with the scintillator oriented in the outward direction (away from the PMT face) was aligned vertically and configured in beam line center. The flat face of the aluminum surface was mounted 3.5~cm above the plane of the beam line and the center of the flat surface was placed 36.3~cm from the west wall of Cave~02, where the beam entered the experimental space. The scatter cells were configured in the plane of the beam line and oriented such that the outward normal of the flat aluminum surfaces faced the target cell. Each of the scatter cells were placed at varying flight paths and angles relative to the center of the target cell, as described in Table~\ref{config}. 

\begin{figure}
\centering
\includegraphics[width=0.9\textwidth]{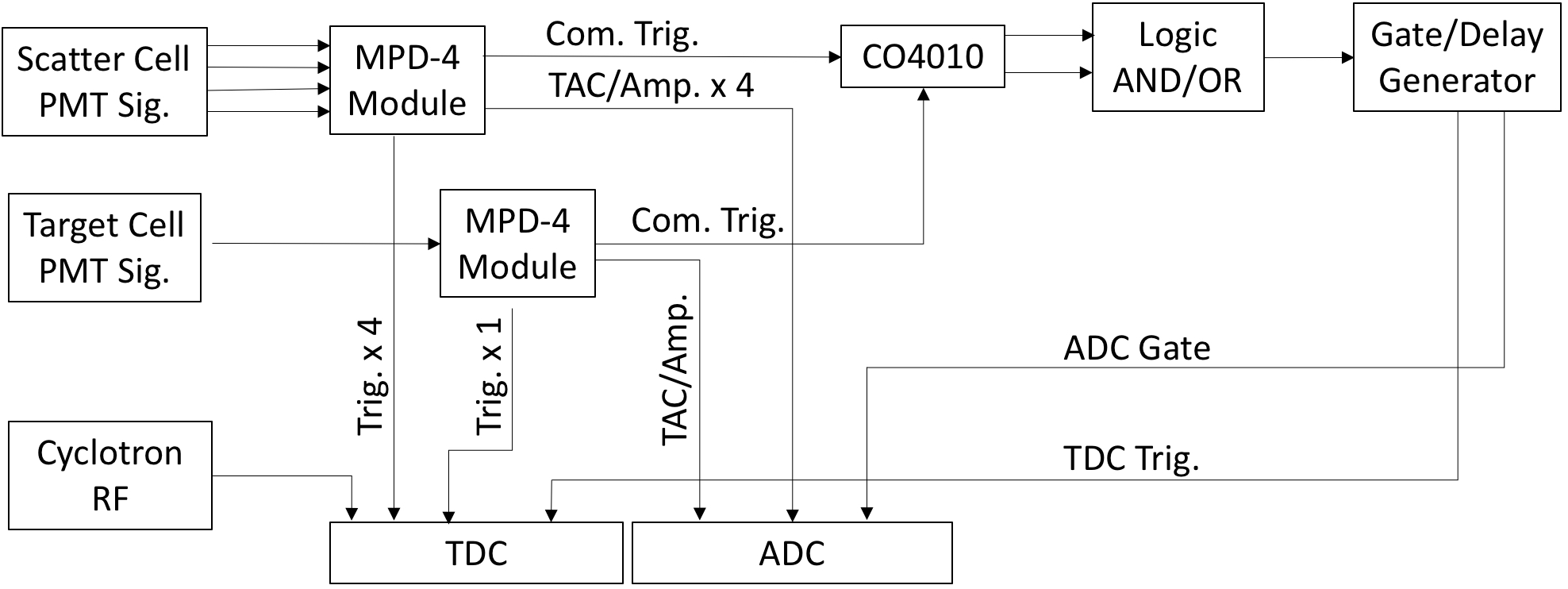}
\caption{Block diagram of the data acquisition circuit. \label{blockdiagram}}
\end{figure}

A block diagram of the signal chain for data acquisition is shown in Fig.~\ref{blockdiagram}. The timing of the PMT output signals was established using constant fraction discrimination (CFD) via two mesytec MPD-4 modules, one to process the four scatter cell signals and one for the target cell signal. The CFD output signals for each cell were fed into a CAEN V1290N multi-hit time-to-digital converter (TDC) operating in trigger matching mode with the clock multiplier of the module set to 195~ps resolution. The pulse height (Amp.) and pulse shape (TAC) data for each input signal were obtained using the MPD-4 modules as fast variable gain input amplifiers whose signal was fed to a CAEN V785 peak sensing analog-to-digital converter (ADC).

The MPD-4 modules produced common trigger gates with a width of 1~$\mu$s that were fed to an ORTEC CO4010 Logic Unit set to produce a 350~ns gate for the target cell and a 10~ns gate for the scatter cells, triggering on the falling edge of the MPD-4 common trigger gate. This allowed for a coincidence window of 350~ns. These signals were fed into a Model 622 LeCroy Quad Coincidence module (labeled `Logic AND/OR' in Fig.~\ref{blockdiagram}), with an AND setting for coincidence measurements and an OR setting for pulse height singles measurements. The resulting signal was then fed into a Model 794 Phillips Scientific Quad Gate/Delay Generator that output a 1~$\mu$s gate, used as the gate input for the ADC. A time-delayed logic signal was used as the TDC trigger. 

The timing of the RF signal from the cyclotron, converted to standard logic pulses using a CFD module fabricated at LBNL, was also recorded using the TDC. Data were acquired for a period of $13.2$~hours for the dTOF measurement and $48.8$~minutes for the target pulse height spectral measurements, and recorded using the MIDAS data acquisition system \cite{MIDAS}. Data reduction was performed using the ROOT data analysis framework \cite{ROOT}.

\section{Analysis}
\label{beam_char}

\subsection{Spectral Shape using Double Time-of-Flight}
\label{shape}

The incoming neutron TOF, $t_{in}$, was defined as the time difference between a cyclotron RF logic pulse and a signal in the target cell. Similarly, the outgoing TOF, $t_{out}$, was defined as the time difference between a signal in the target cell and a signal in one of the scatter cells. The TAC pulse shape signals from the ADC were histogrammed and the neutron and gamma contributions were fit to a superposition of Gaussian distributions. Neutron events were accepted within three standard deviations of the mean of the neutron feature to account for 99.7\% of the neutron events. While this high neutron acceptance rate increases the gamma contribution at low pulse heights, this background contribution is accounted for in the analysis that follows. Timing calibrations for the target and scatter cell signals were performed using photon events and the known flight paths, shown in Table~\ref{config}. 

Using neutron-proton elastic scattering kinematics, the expected incoming neutron TOF, $t_{calc}$, was also calculated in terms of the measured outgoing neutron TOF, $t_{out}$, and the neutron scattering angle, $\phi$:
\begin{equation}
\label{en-calc}
E_n = \frac{E_{n^{\prime}}}{\cos^2\phi},
\end{equation}
where $\phi$ is measured relative to beam line center. The kinetic energy of the incoming neutron, $E_n$, and outgoing neutron, $E_{n^{\prime}}$, is related to the TOF through the relativistic energy-time relation:
\begin{equation}
\label{energy-time}
E_n = \Bigg(\frac{1}{\sqrt{1-(\frac{(L/t)}{c})^2}}-1\Bigg) m_n c^2.
\end{equation}
Here, $m_n$ is the mass of the neutron, $c$ is the speed of light, and $L$ is the relevant flight path.

\begin{figure}
\centering
\includegraphics[width=0.9\textwidth]{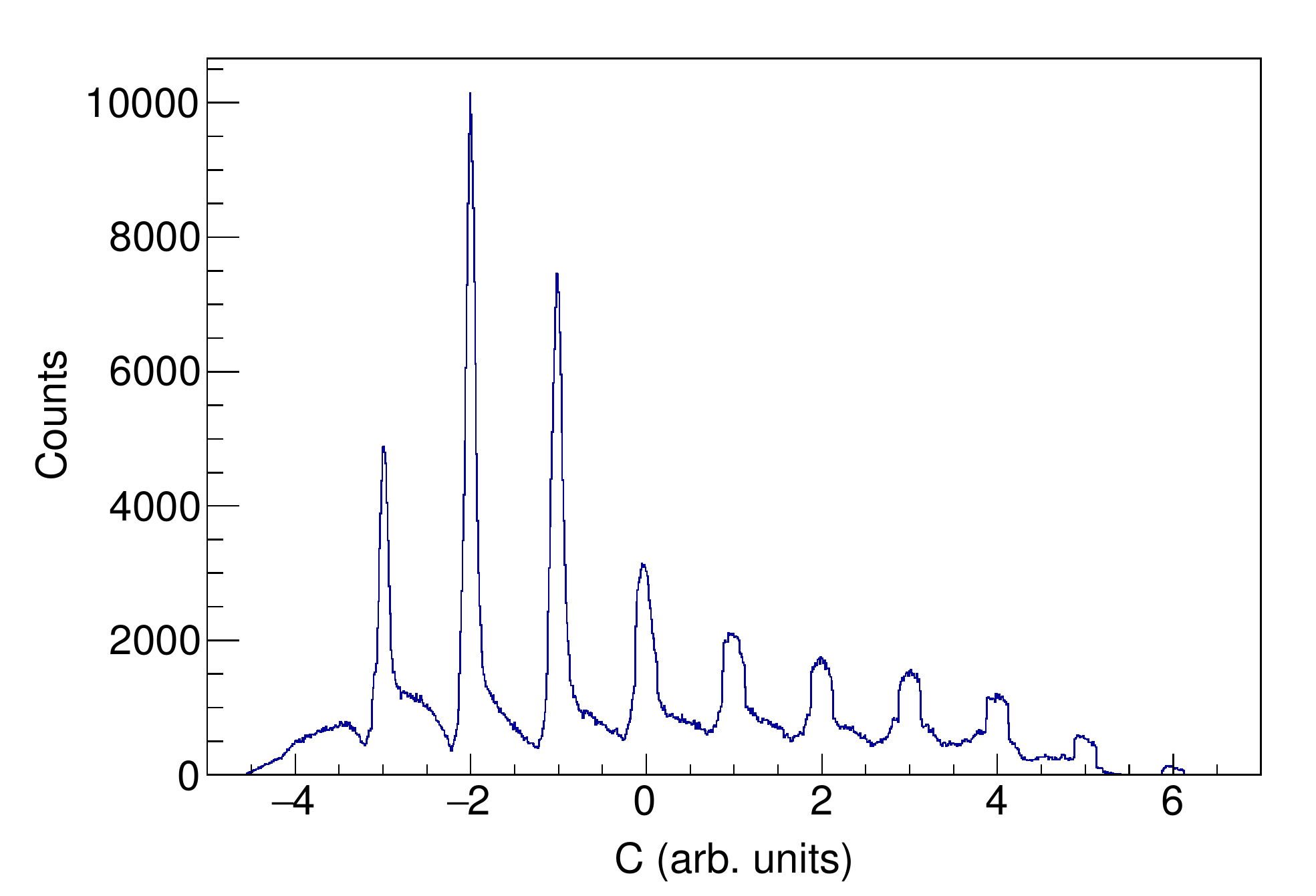}
\caption{Distribution of the values of $C$, a constant reflective of the number of cyclotron beam pulse offsets, used to assign the reconstructed incoming TOF. \label{Cdist}}
\end{figure}

An event reconstruction algorithm was used to disambiguate neutrons originating from different cyclotron RF cycles. This was accomplished by identifying an integer number of cyclotron RF cycles by which the incoming TOF signal was offset. On an event-by-event basis, a constant, $C$, reflective of the number of cyclotron pulse offsets, was calculated as the difference between the measured incoming TOF, $t_{in}$, and the calculated incoming TOF, $t_{calc}$, divided by the cyclotron period, $T$: 
\begin{equation}
C = \frac{t_{in}-t_{calc}}{T}.
\end{equation}
The following condition was applied:
\begin{equation}
\label{recon-gate}
| T*(C - \text{Round}(C)) | < 20~\text{ns}, 
\end{equation}
where the time difference between the measured and calculated incoming TOF was less than 20~ns for a given beam pulse offset. Events satisfying this condition were assigned a reconstructed TOF value, $t_{r}$:
\begin{equation}
\label{recon}
t_r = t_{in} +T*\text{Round}(C),
\end{equation}
thereby associating each temporally overlapped event with the appropriate cyclotron pulse. A distribution of the $C$ values for this measurement is shown in Fig.~\ref{Cdist}. 

Conservative lower detection thresholds for the target and scatter cells and upper detection thresholds for the scatter cells were established in post-processing. To characterize the energy equivalent of the upper and lower detection thresholds for the scatter cells, pulse height calibration was performed by comparing the measured pulse height spectra for $^{137}$Cs, $^{22}$Na, and $^{60}$Co with a GEANT4 simulated pulse height distribution folded with the energy-dependent detector resolution \cite{dietze}. From this comparison, linear pulse height calibration parameters were determined for each target and scatter cell. The upper and lower pulse height threshold values were converted to light yield in MeV electron equivalent (MeVee), and then to proton energy deposited using a numerical integration of Birks relation \cite{birks}:
\begin{equation}
\label{birks}
\frac{dL}{dE}=\frac{S}{1+kB(dE/dx)},
\end{equation}
where $dL/dE$ is the scintillation light yield per unit energy loss and $dE/dx$ is the energy-dependent stopping power of a proton in the detection medium. The scintillation efficiency, $S$, and Birks parameter, $kB$, were obtained from an empirical fit of the light yield of EJ-309 as a function of proton energy deposited in the target cell, where $S$ was determined to be $0.8394 \pm 0.0726$~MeVee/MeV and $kB$ was $6.994 \pm 1.196$ mg/(MeV cm$^2$).\footnote{Due to the 20-ns pulse height integration window of the mesytec MPD-4 modules, the light yield parameters cited here may not reflect the fundamental characteristics of the scintillating medium \cite{ruben}.} Stopping power values were taken from the SRIM library \cite{SRIM}. The upper and lower threshold values in MeVee are provided in Table~\ref{thresh}.

\begin{table}
\centering
\begin{tabular}{ccccc}
\hline
Detector & Lwr TH (MeVee) & Up TH (MeVee) \\
\hline
Target    & $0.0601 \pm 0.0012$ &  N/A   \\
Scatter 0 & $0.0414 \pm 0.0006$ & $1.0271 \pm 0.0007$ \\ 
Scatter 1 & $0.0489 \pm 0.0007$ & $1.1290 \pm 0.0027$ \\
Scatter 2 & $0.0531 \pm 0.0005$ & $0.9548 \pm 0.0006$ \\
Scatter 3 & $0.0469 \pm 0.0005$ & $1.0611 \pm 0.0006$ \\
\hline
 \end{tabular}
 \caption{The lower (Lwr TH) and upper (Up TH) detection thresholds in MeV electron equivalent (MeVee) and associated uncertainty for the target and scatter cells in the neutron detection array. \label{thresh}}
 \end{table}

\begin{figure*}
\centering
\includegraphics[width=1.0\textwidth]{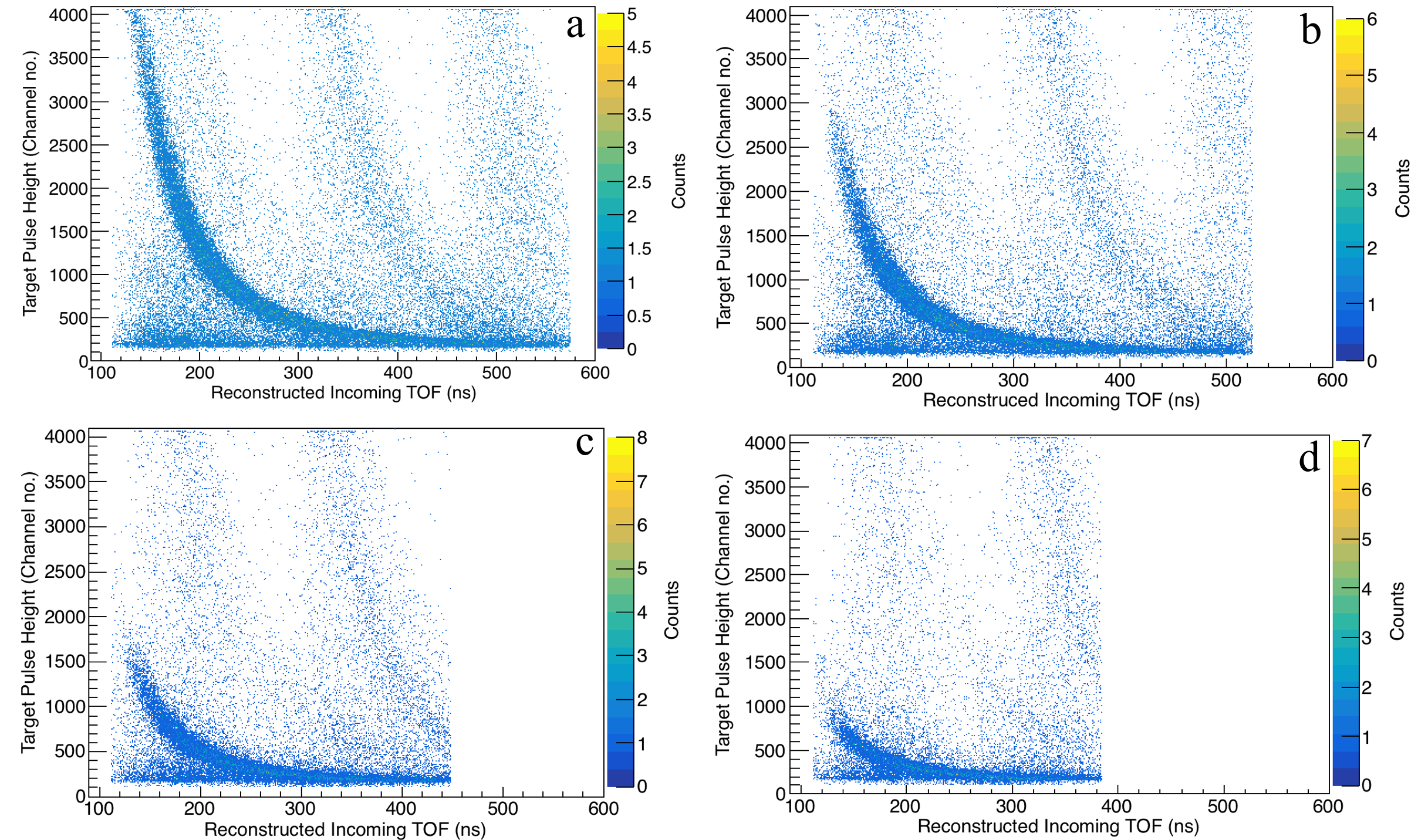}
\caption{Pulse height in the target cell as a function of the reconstructed incoming TOF for the four angles corresponding to the positions of (a) Scatter Cell 0, (b) Scatter Cell 1, (c) Scatter Cell 2 and (d) Scatter Cell 3.\label{phvtof}}
\end{figure*}

\begin{figure*}
\centering
\includegraphics[width=1.0\textwidth]{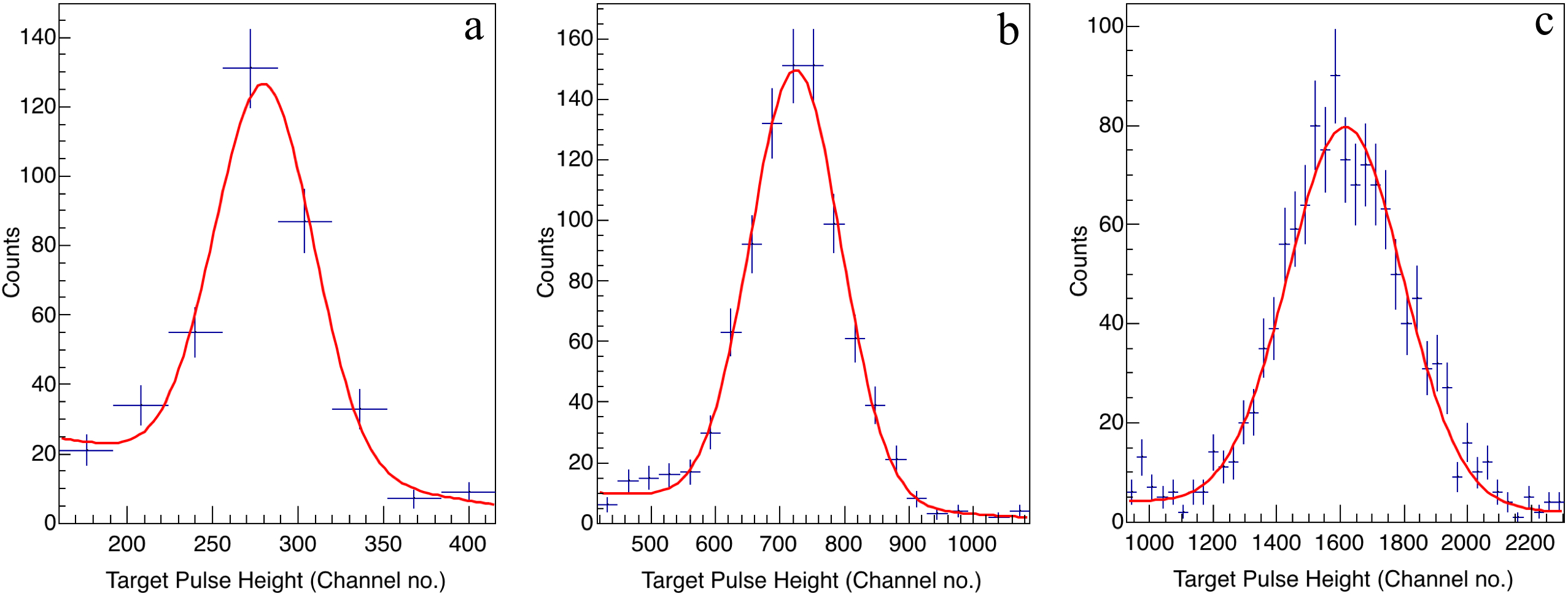}
\caption{Neutron-proton elastic scattering features as a function of pulse height for three different neutron energy ranges in $6.12$-ns time bins at (a) $382.46$~ns ($1.67$~MeV), (b) $247.80$~ns ($4.00$~MeV) and (c) $186.58$~ns ($7.09$~MeV) .\label{slices}}
\end{figure*}

A neutron-gated histogram of the target pulse height as a function of reconstructed incoming TOF with scatter cell thresholds applied is shown in Fig.~\ref{phvtof} for the four scatter cells. Projections of the pulse height spectra were obtained in $6.12$-ns bins with representative examples for three different energy ranges shown in Fig.~\ref{slices}. The peaks in the pulse height projections were fit using a Gaussian distribution and linear background term to determine the number of neutron-proton elastic scattering events as a function of neutron energy.

\subsection{Efficiency Simulation}
\label{eff-sec}

Simulation of the efficiency of the detector array was performed using both the MCNP6 \cite{mcnp6} and GEANT4 \cite{geant4} software packages. The simulations employed a square parallel uniform source of 0-25~MeV neutrons directly incident on an overfilled target cell with four scatter cells positioned in an air environment in the geometry described in Table~\ref{config}. The cells were a direct reflection of those employed in the experiment, composed of EJ-309 in an aluminum housing. Processing of the MCNP6 simulation output defined a ``good" event as a single neutron-proton scattering event in the target cell with a proton recoil energy above the lower target cell threshold, and one or more scattering events in a given scatter cell. The scattered neutron was further subjected to the following constraints: no collisions between the target cell and the scatter cells, and the sum of the proton energy depositions in the scatter cell was within the upper and lower scatter cell threshold values. 

For the GEANT4 simulation, the PMT assembly, including the acrylic window between the glass of the PMT and the active volume, the borosilicate glass, and the magnetic shield surrounding the PMT, were also modeled. A ``good" event required a single neutron-proton scattering event in the target cell that deposited more energy than the lower target cell threshold and one or more neutron interactions in a given scatter cell that resulted in an energy deposition within the upper and lower detection thresholds in that scatter cell. In post-processing, the simulated TOF was randomized with the uncertainty in the experimentally measured incoming (6.12~ns) and outgoing (1.12~ns) TOF. The TOF between the target and scatter cells was then required to be less than 169.71~ns, consistent with the constraint applied in the experimental data analysis. Further, the simulated incoming neutron TOF and expected incoming TOF (calculated from the simulated outgoing TOF and the neutron scattering angle) was required to be within a 20~ns window as described in Eq.~\ref{recon-gate} to reflect the experimental data reduction. 

\begin{figure}
\centering
\includegraphics[width=0.9\textwidth]{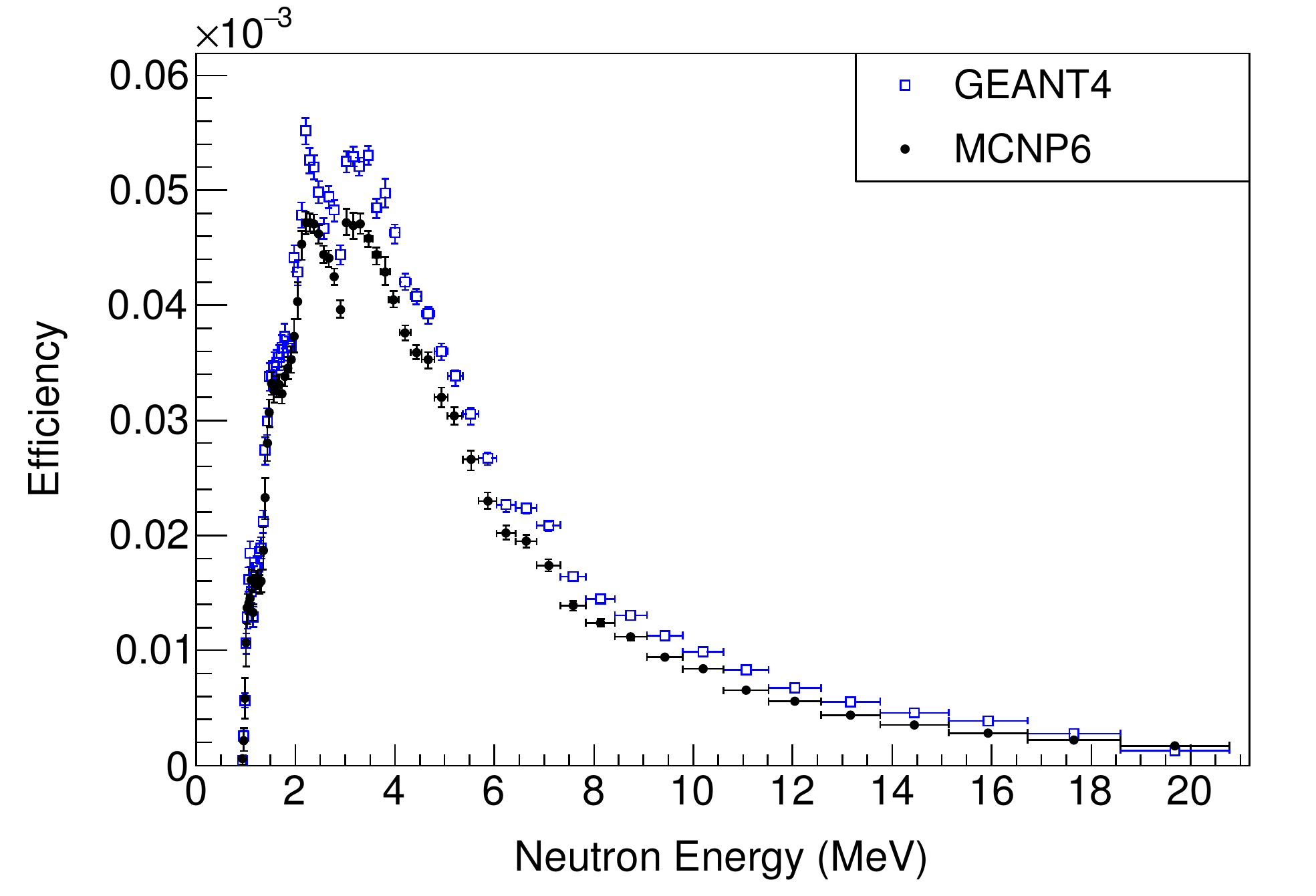}
\caption{Detection efficiency simulated using the MCNP6 and GEANT4 software packages. The closed circles (black) and open squares (blue) represent the energy-dependent neutron detection efficiency obtained from MCNP6 and GEANT4, respectively. \label{eff}}
\end{figure}

A plot of the efficiency of the neutron detection array simulated with GEANT4 and MCNP6 is shown in Fig.~\ref{eff}. The two simulation outputs agree within estimated uncertainty from $0-2$~MeV. Above 2~MeV, the results are discrepant but display similar features consistent with the literature \cite{lone,brede}. The disagreement between the two simulation outputs is due in part to the differences in the simulation environments and post-processing. The GEANT4 simulation included uncertainty on the TOF, allowed neutron interactions between the target and scatter cells for events that fell within the scattered TOF window, and post-processing incorporated the event reconstruction algorithm and associated data reduction procedures. Both simulation outputs were used to correct the spectral shape obtained in Sec.~\ref{shape} while foil activation analysis was used for an absolute normalization of the energy differential neutron spectra, as described in Sec.~\ref{foils}. The spectra obtained using the two different simulation platforms are compared in Sec.~\ref{results-yo} and the results exhibit agreement within error over the full energy range of the measurement.

The total uncertainty on the efficiency simulation output included both a statistical component and an asymmetric systematic component resulting from the detection threshold values. The asymmetric systematic contribution was determined by performing three different determinations of the efficiency. The first output the nominal efficiency using the threshold values shown in Table~\ref{thresh}. The second calculation output the upper bound on the efficiency, which used the threshold values in Table~\ref{thresh} minus the uncertainty on these values for the lower thresholds and plus the uncertainty on these values for the upper thresholds. The third calculation output the lower bound on the efficiency, which used the lower thresholds plus the uncertainty and the upper thresholds minus the uncertainty. The upper bound of the uncertainty was taken to be the quadrature sum of the difference between the upper limit on the efficiency and the nominal efficiency and the statistical error on the nominal efficiency. The lower bound of the uncertainty was determined using the same method.

\subsection{Foil Activation Analysis}
\label{foils}

To correct for dead time effects in the data acquisition, the absolute number of neutrons incident on the target scintillator was determined using foil activation analysis. Two foils, aluminum and indium, were placed just before the target cell in-line with the neutron beam and irradiated at $\sim$1.8 $\mu$A for 4~hours and 47~minutes. The aluminum and indium foils were both 5 mm in diameter and 1~mm thick with a weight of 5.36~g and 14.34~g, respectively. The Al foil was composed of of 99.9~\% $^{27}${Al} and the In foil was composed of 95.71~\% $^{115}${In} and 4.29~\% $^{113}${In}. The $^{27}${Al}(n,$\alpha$)$^{24}${Na} and the $^{115}${In}(n,n$^{\prime}$)$^{115m}${In} reactions were used to produce neutron activation products. 

\begin{figure}
\centering
\includegraphics[width=0.9\textwidth]{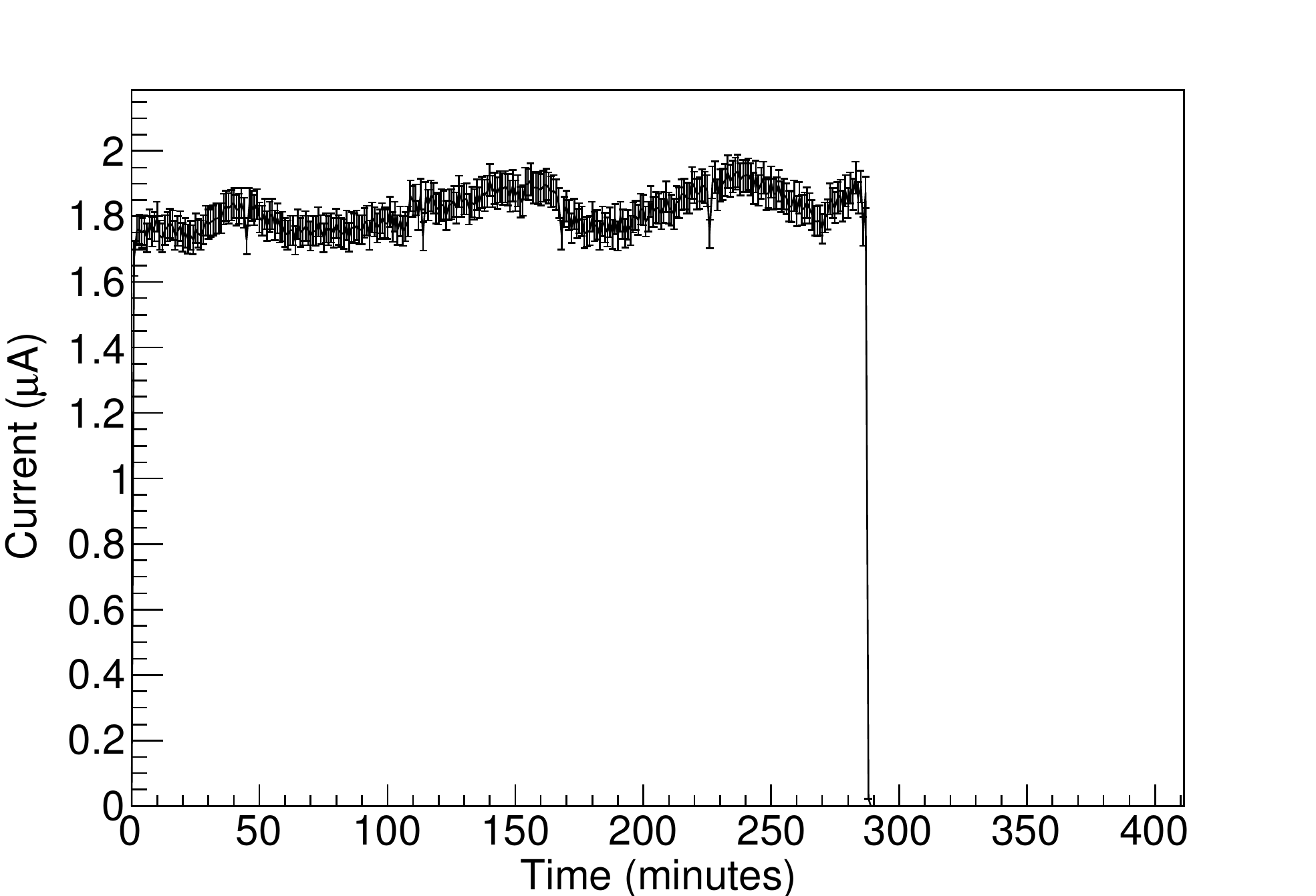}
\caption{Beam current in microamperes during the foil irradiation and resting period.\label{current}}
\end{figure}

The number of activation products produced, $N$, is described by the following differential equation:
\begin{equation}
\label{foilEQ}
\frac{dN(t)}{dt} =  N_{T} \langle \sigma \rangle \phi(t)-N \lambda, 
\end{equation}
where $N_{T}$ is the number of $^{27}${Al} or $^{115}${In} nuclei in the foils at the start of irradiation and $\lambda$ is the decay constant. The effective activation cross section, $\langle \sigma \rangle$, is given by:
\begin{equation}
\label{eff-sig}
\langle \sigma \rangle = \frac{\int \Phi(E) \sigma(E) dE}{\int \Phi(E) dE},
\end{equation}
where $\Phi(E)$ is the energy-differential neutron spectrum and $\sigma(E)$ is the energy-dependent cross section for the activation reactions. The time-dependent neutron flux, $\phi(t)$, can be written as:
\begin{equation}
\label{flux2current}
\phi(t)=\kappa I(t),
\end{equation}
where $\kappa$ is a normalization constant and $I(t)$ is the time-dependent deuteron beam current, which was measured using a Model 6060 Pulse Link Extended Range Neutron Area Monitor and normalized to the output of a current integrator. The neutron monitor was located on the east wall of the experimental space and provided a signal proportional to the number of neutron interactions on a per minute basis. The current integrator measured the total charge accumulated over a given period of time on a Faraday cup located at the breakup target. A reading from the current integrator was taken every thirty minutes and divided by the time duration of the charge accumulation period. The normalized neutron monitor data were used to determine the relative change in the beam current on a minute-by-minute basis. A plot of the beam current as a function of time during the irradiation period is shown in Fig.~\ref{current}.

Equation~\ref{foilEQ} can be solved using the boundary conditions $N(0)=0$ and $N(t_{c}) = N_{c}$, where $t_{c}$ is the start time of the counting period and $N_c$ is the number of activation products at the start of the counting period. This gives:
\begin{equation}
\label{foilEQ_sol}
N_c =  N_{T} \langle \sigma \rangle \kappa e^{-\lambda t_{c}} \int_{0}^{t_{c}} I(t) e^{-\lambda t} dt.
\end{equation} 
Here, $\int_{0}^{t_{c}} I(t) e^{-\lambda t} dt$ is obtained using numerical integration. The law of radioactive decay allows for the determination of the number of activation products that decayed over the counting period:
\begin{equation}
\label{raddecay}
\frac{dN(t)}{dt} = -N \lambda.
\end{equation}
Integrating Eq.~\ref{raddecay} over the period of time from the start of the counting period, $t_{c}$, to the end time, $t_{s}$, with the boundary conditions $N(t_{c}) = N_{c}$ and $N(t_s)= N_{s}$, where the number of decays during the counting period is $N_{d} = N_c - N_s$, produces the following:
\begin{equation}
\label{Ndecays}
N_{d} = N_c (1-e^{-\lambda(t_s - t_c)}).
\end{equation}
Solving for $N_c$ and substituting this into Eq.~\ref{foilEQ_sol} yields the following expression for $\kappa$:
\begin{equation}
\label{kappa}
\kappa = \frac{N_{d}e^{\lambda t_{c}}}{(1-e^{-\lambda(t_s - t_c)})N_{T} \langle \sigma \rangle \int_{0}^{t_{c}} I(t) e^{-\lambda t} dt}.
\end{equation}

The measurement of the number of decays, $N_d$, was performed using a coaxial high-purity Germanium (HPGe) detector with a diameter of 64.9~mm, length of 57.8~mm, and a 0.5-mm-thick beryllium window on the front face of the detector. The detector has a 46.8\% relative efficiency and was oriented in an upward facing direction surrounded by a lead casing. An ORTEC ASPEC-927 multichannel analyzer with two 14-bit ADCs was used to collect data and interface with MAESTRO software \cite{maestro}. The 336.2~keV and 1368.4~keV $\gamma$-ray transitions in $^{115m}${In} and $^{24}${Na}, respectively, were used to determine the number of decays observed in the HPGe detector. Several calibration sources with known activity--$^{152}$Eu, $^{60}$Co, $^{137}$Cs, and $^{133}$Ba--were used to perform energy and efficiency calibrations of the HPGe detector. The number of measured decays was adjusted using the energy-dependent efficiency of the HPGe detector and the branching ratio of the $\gamma$-ray transitions to provide the number of decays, $N_d$. The normalization constant, $\kappa$ has units of cm$^{-2}$ C$^{-1}$ and can be converted to units of sr$^{-1}$ $\mu$C$^{-1}$ and multiplied by the unit-normalized energy-differential neutron spectrum to provide the energy-differential neutron flux.  

\section{Results and Discussion}
\label{results-yo}

\begin{figure}
\centering
\includegraphics[width=0.9\textwidth]{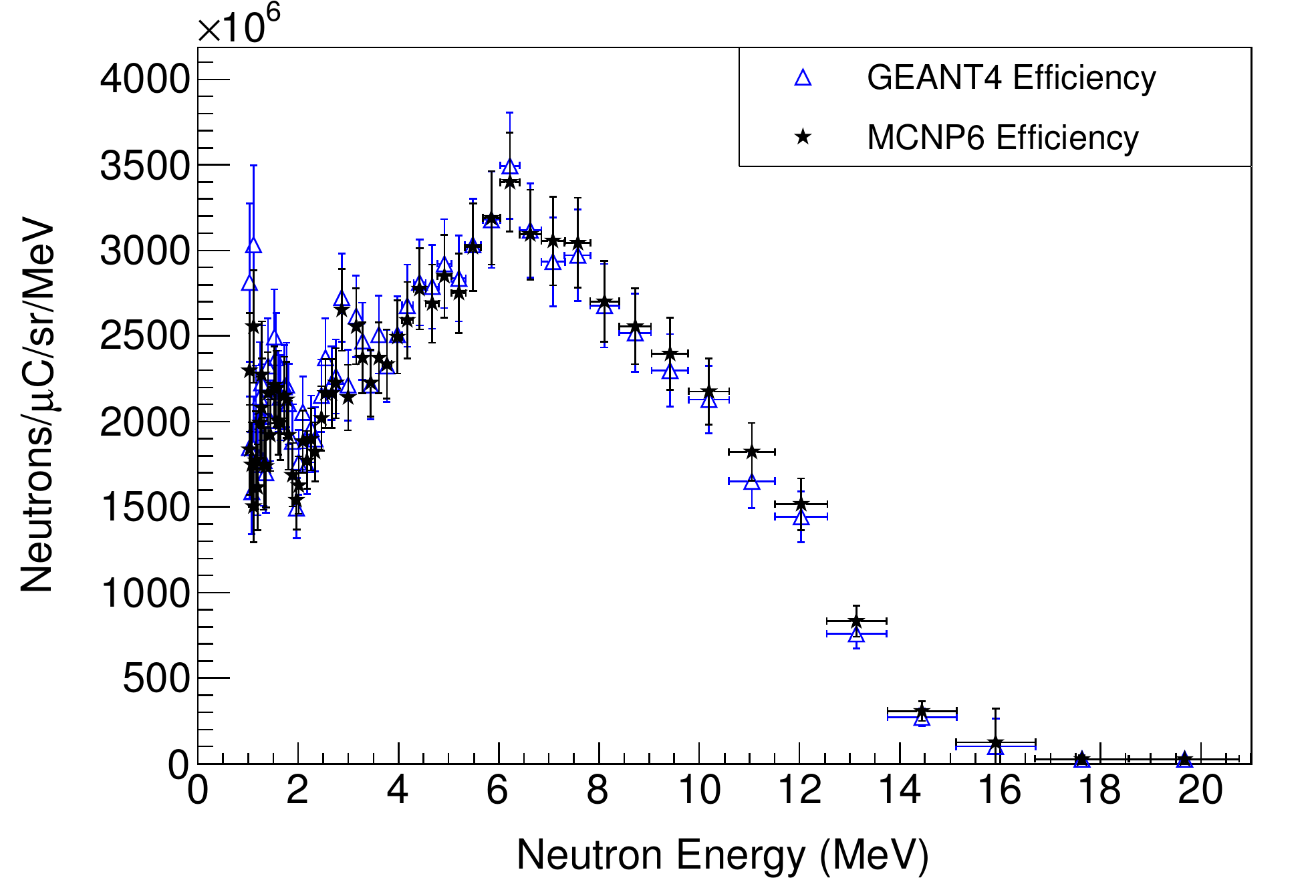}
\caption{Efficiency-corrected and In-foil-normalized energy differential neutron spectra using system detection efficiencies obtained via two different platforms. The spectra shown with open triangles (blue) and closed stars (black) use the system detection efficiency simulated by GEANT4 and MCNP6, respectively. The error bars represent both the statistical and systematic uncertainty. \label{infoil}}
\end{figure}

\begin{figure}
\centering
\includegraphics[width=0.9\textwidth]{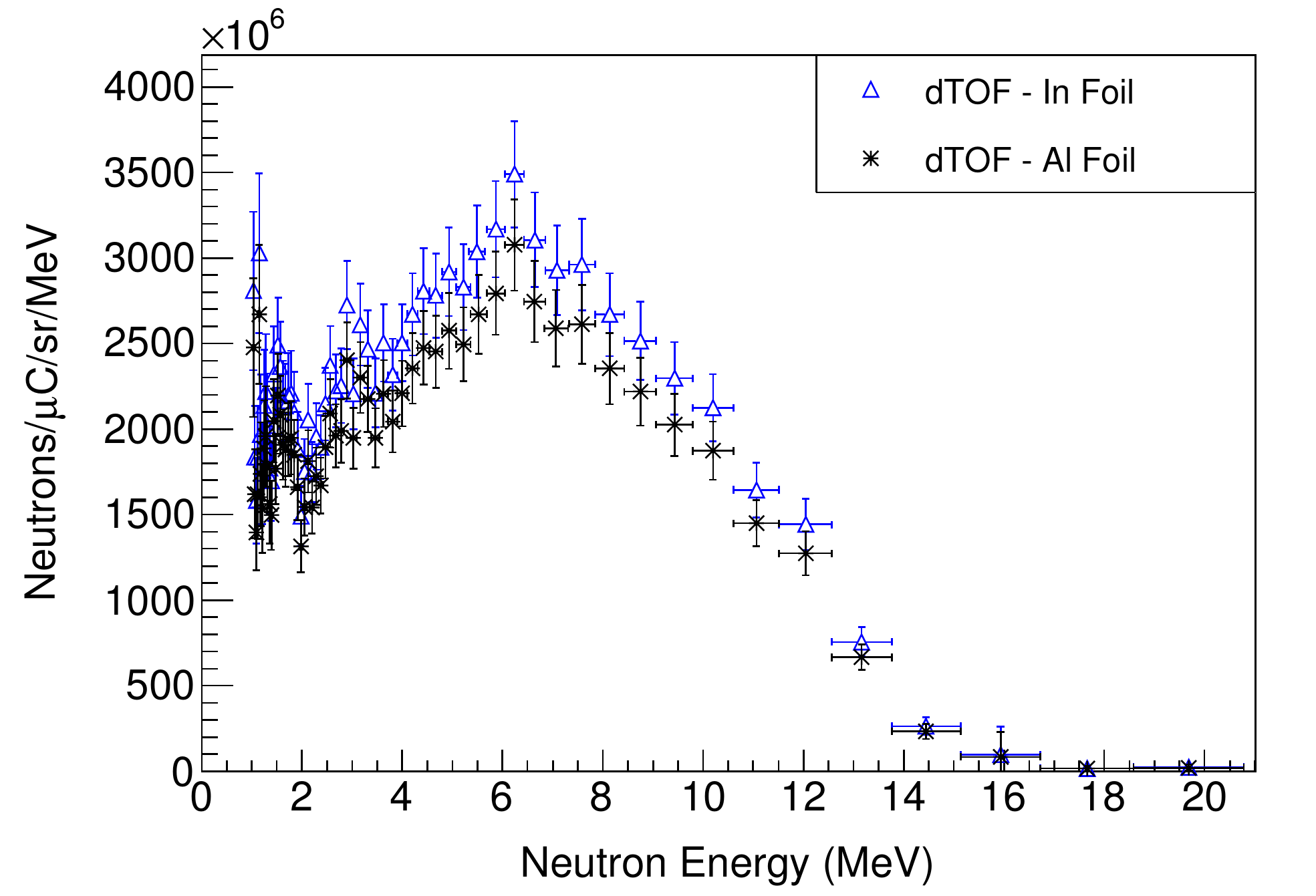}
\caption{Efficiency-corrected and foil-normalized energy differential neutron spectrum for In and Al activation foils shown with open triangles (blue) and asterisks (black), respectively. The error bars represent both the statistical and systematic uncertainty. \label{foil}}
\end{figure}

The spectral shape obtained in Sec.~\ref{shape} was corrected using the detection efficiency obtained from both MCNP6 and GEANT4 simulations and normalized using the In activation foil to provide the energy-differential neutron spectra shown in Fig.~\ref{infoil}. Despite the disparities in the system efficiency determined by MCNP6 and GEANT4 outlined in Sec.~\ref{eff-sec}, the spectra agree within error over the full energy range of the measurement. The energy-differential neutron spectrum was also determined using the GEANT4 detection efficiency and the In and Al foil activation analyses respectively, as shown in Fig.~\ref{foil}. The results from normalization using the two different foils agree within error over the full energy range of the measurement. The integral of the spectra from 1.0~MeV to 20.0~MeV agree within one sigma of the uncertainty on the measurements, as shown in Table~\ref{integral}. The dTOF neutron spectral data are provided in tabular form online as supplementary material to this article. This includes the energy-differential neutron flux data with efficiency corrections obtained from both the MCNP and GEANT simulations using the In foil normalization, as it has a lower activation threshold of 336~keV and samples the full energy range of the measurement.

\begin{table*}
\centering
\begin{tabular}{cccc}
\hline
   & Integrated Neutron Flux ($\frac{\mathrm{neutrons}}{\mathrm{\mu C sr}}$) & Uncertainty ($\frac{\mathrm{neutrons}}{\mathrm{\mu C sr}}$)  \\
\hline
In Activation  & $2.913 \times10^{10}$ & $+  2.462 \times10^{9}$/$- 2.441 \times10^{9}$ \\
Al Activation & $2.570 \times10^{10}$ & $+ 2.129 \times10^{9}$/$- 2.083 \times10^9$ \\
HEPROW Unfolded & $2.875 \times10^{10}$ & $+ 2.815 \times10^8$/$- 2.815 \times10^8$\\
GRAVEL Unfolded & $2.884 \times10^{10}$ & $+ 4.733 \times10^8$/$- 4.733 \times10^8$\\ 
\hline
 \end{tabular}
 \caption{Comparison of the integrated flux from 1.0~MeV to 20.0~MeV for the In and Al-normalized dTOF spectra, the HEPROW unfolded spectrum, and the GRAVEL unfolded spectrum. \label{integral}}
 \end{table*}

The error bars for the data shown in Figs.~\ref{infoil} and~\ref{foil} represent both the statistical and systematic uncertainty. The systematic uncertainty was obtained using a Monte Carlo approach. A normal distribution about zero was created for each parameter used to calculate the $\kappa$ value in Eq.~\ref{kappa}, where the standard deviation of each distribution was set to the uncertainty on the parameters. These values were then randomly sampled using 10,000 trials and added to the nominal parameter value in Eq.~\ref{kappa}. The $\kappa$ value was then calculated for each trial and the resulting $\kappa$ distribution was fit with a Gaussian function. The centroid of the distribution was taken as the $\kappa$ value and the standard deviation of the distribution was used to represent the uncertainty on $\kappa$. The asymmetric uncertainty of the simulated efficiency, due to the uncertainty in the upper and lower detection thresholds, was accounted for by performing the foil normalization procedure outlined in Sec.~\ref{foils} for each of the three efficiency simulations described in Sec.~\ref{eff-sec}. The uncertainty from the Monte Carlo and the statistical uncertainty on the spectral measurement were then added in quadrature with the difference between the upper, nominal, and lower bounds resulting from the foil normalization output, to yield asymmetric error bars on the final spectra.

\begin{figure}
\centering
\includegraphics[width=0.9\textwidth]{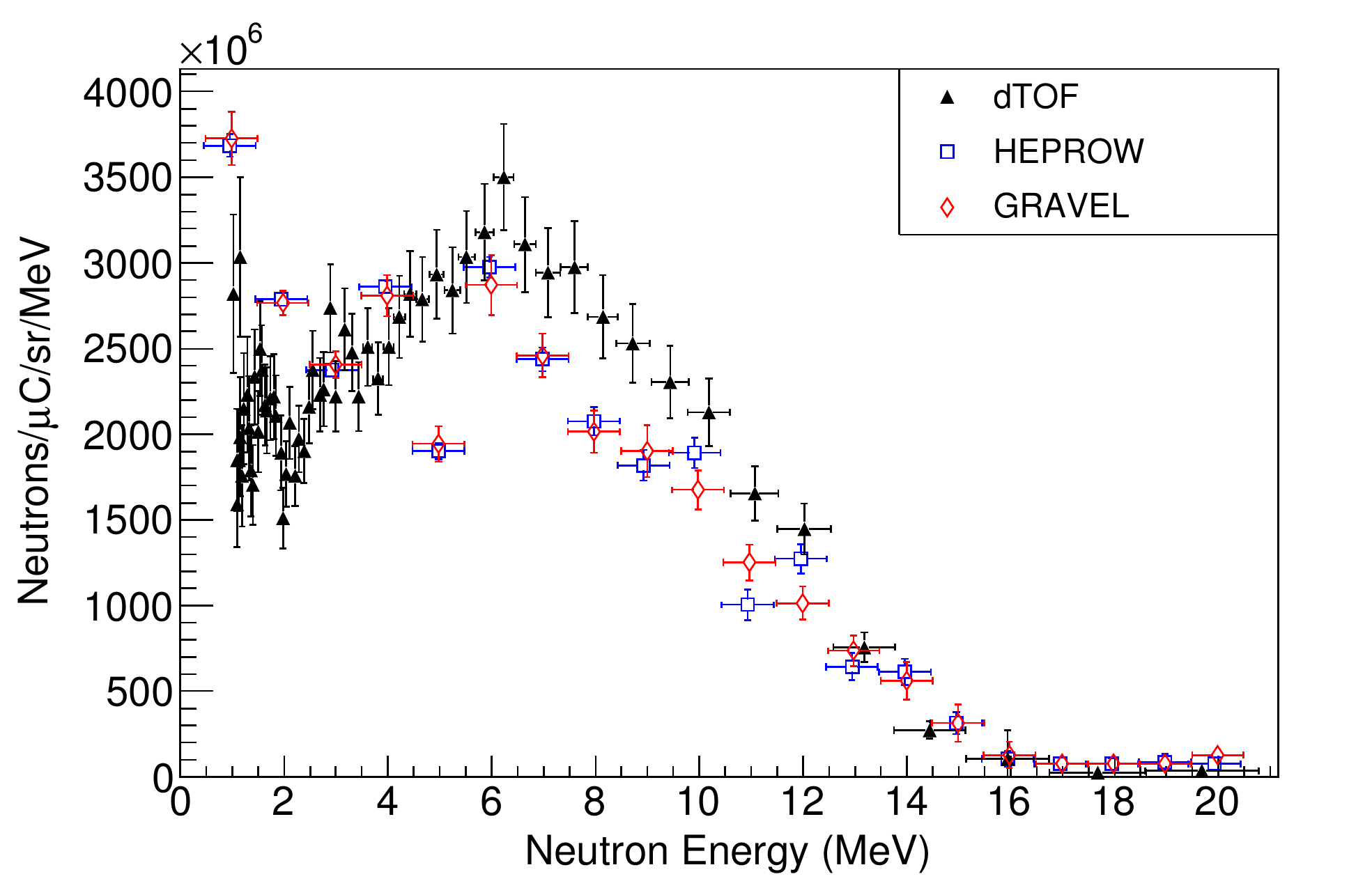}
\caption{Efficiency-corrected and In-foil-normalized energy differential dTOF neutron spectrum compared to spectra obtained using the GRAVEL and HEPROW pulse height spectrum unfolding algorithms. \label{all}}
\end{figure}

A comparison of the GEANT4 efficiency-corrected dTOF spectrum obtained using the In foil normalization and neutron spectra obtained using pulse height spectrum unfolding is shown in Fig.~\ref{all}. Two established neutron pulse height spectrum unfolding techniques were used to obtain the unfolded spectra: HEPROW \cite{heprow} and GRAVEL \cite{gravel}. The HEPROW spectrum unfolding was performed using both the GRAVEL and MIEKE sequences from the HEPROW package, and the GRAVEL spectrum was obtained using an independent GRAVEL unfolding algorithm. No additional a priori or regularization constraints (i.e., no assumptions on the smoothness of the spectra) were applied. Using the neutron singles pulse height measurements from the target scintillator shown in Fig.~\ref{layout}, pulse shape discrimination was accomplished via a pulse-height-dependent gate on the neutron feature of the pulse shape (TAC) output from the MPD-4 module described in Sec.~\ref{exp}. A lower detection threshold of 0.234~MeVee was applied to the neutron-gated pulse height spectrum. The characteristic response matrix of the EJ-309 detector was simulated using an experimentally-benchmarked GEANT4 simulation and smeared with a measured resolution function \cite{dietze,geant4}. The response matrix was normalized using the detection efficiency obtained from the GEANT4 simulation. The unfolded spectra were normalized for solid angle, integrated current delivered (using the beam current measured at the Faraday cup on the breakup target as described in Sec.~\ref{beam_char}), and corrected for detection system dead time.

\begin{figure}
\centering
\includegraphics[width=0.9\textwidth]{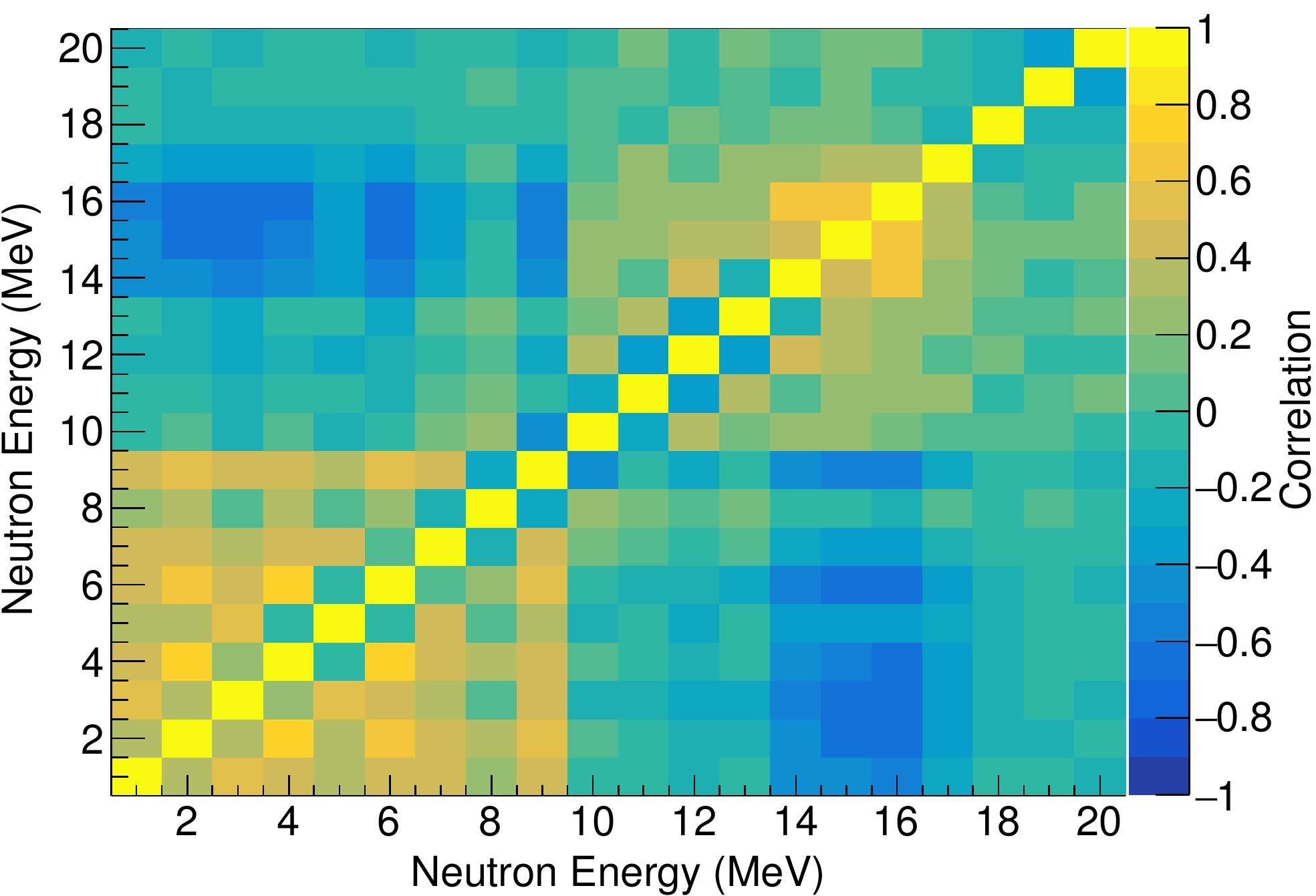}
\caption{Correlation matrix between energy bins for the neutron spectrum obtained using the GRAVEL pulse height spectrum unfolding algorithm. The color bar represents the correlation, ranging from -1.0 to 1.0. \label{corr}}
\end{figure}

For the HEPROW result, MIEKE was used for uncertainty quantification through an assessment of the probability density of the unfolded GRAVEL neutron spectrum using Bayesian and Maximum Entropy inference methods and Monte Carlo sampling to inform expectation values. These methods propagate statistical uncertainty from the pulse height spectrum measurement and ambiguity (i.e. uncertainty due to degeneracy of the solution), but do not take into account uncertainties on the light yield relation parameters. For the independent GRAVEL algorithm, the uncertainty was quantified via a Monte Carlo sampling of the input parameters about their associated uncertainty. These include the statistical error on the pulse height spectrum and the response matrix, as well as systematic contributions from the uncertainty on the MeVee calibration parameters and the light yield relation parameters. The Monte Carlo of the GRAVEL solution was used to generate a correlation plot between different energy bins for the GRAVEL algorithm output as shown in Fig.~\ref{corr}. The linear correlation between energy bins for the resulting neutron distributions was calculated, where any two bins were treated as random variables in a Pearson correlation.

While the dTOF spectrum and the HEPROW and GRAVEL results are discrepant in shape, the integrated flux obtained using the different approaches agree within two standard deviations. The shape of the dTOF spectrum more closely resembles that for d-breakup spectra on similar breakup targets in the literature \cite{meulders,lone,brede}. The discrepancies between the dTOF and unfolded spectra may be due in part to lack of regularization constraints in the unfolding algorithms leading to shape fluctuations and a limitation of the spectrum unfolding technique given the degeneracy of the response functions. The latter is demonstrated in Fig.~\ref{corr}, where the highly negative correlation coefficients in the discrepant energy ranges of the GRAVEL neutron spectrum suggest degenerate components to the response matrix used as an input to both the GRAVEL and HEPROW spectrum unfolding algorithms. Similar shapes of the response functions in the energy ranges of $\sim$2-4~MeV/6~MeV/9~MeV and $14-16$~MeV suggest that these energy bins are not completely orthogonal. From trial to trial, the energy bins corresponding to the more degenerate response functions tend to fluctuate as the GRAVEL solution samples many valid minima. This effect can be seen as a strong anticorrelation in the correlation matrix.  

A shelf at $16$~MeV is evident in the neutron spectra obtained from dTOF, HEPROW and GRAVEL. This is likely a result of neutrons produced in the $^{9}$Be(d,n)$^{10}$Be reaction, where the $^{10}$Be nucleus is left in the ground state or an excited state and has been observed in previous results \cite{weaver,lone,brede,lone2}. Table~\ref{integral} provides a comparison of the integral values from 1.0~MeV to 20.0~MeV of the dTOF, HEPROW and GRAVEL spectra. 


\section{Summary}
\label{conc}

A dTOF method for neutron spectroscopy with pulsed broad spectrum sources has been demonstrated. Using an over-constrained system and neutron-proton elastic scattering kinematics, neutrons from temporally overlapped frames were associated with the appropriate beam pulse at the 88-Inch Cyclotron at LBNL. The energy differential thick-target 16~MeV deuteron breakup neutron spectrum was measured using this technique and compared to results obtained from HEPROW and GRAVEL pulse height spectrum unfolding algorithms. The spectra are discrepant in shape, possibly due to degeneracies in the response matrix, but the integral neutron yield from 1.0~MeV to 20.0~MeV agrees within two standard deviations for the dTOF and unfolded results. The dTOF method provides a new capability in neutron spectroscopy that eliminates ambiguities associated with ``wrap-around" neutrons and expands opportunities for fundamental and applied physics measurements using pulsed broad spectrum neutron sources. 


\section*{Acknowledgments}

The authors gratefully acknowledge the 88-Inch Cyclotron operations and facilities staff for their help in performing these experiments and N.M. Brickner for graphic design support. This material is based upon work supported by the Department of Energy National Nuclear Security Administration through the Nuclear Science and Security Consortium [Award Nos. DE-NA0000979, DE-NA0003180], the University of California Office of the President [Award No. 12-LR-238745], the National Science Foundation Graduate Research Fellowship [Grant No. NSF 11-582], and performed under the auspices of the U.S. Department of Energy by Lawrence Livermore National Laboratory [Contract DE-AC52-07NA27344] and by Lawrence Berkeley National Laboratory [Contract No. DE-AC02-05CH11231].

\section*{References}

\bibliography{./Harrig-Preprint}
\bibliographystyle{unsrtnat}


\end{document}